\documentclass[conference]{IEEEtran}
\IEEEoverridecommandlockouts
\usepackage{cite}
\usepackage{amsmath,amssymb,amsfonts}
\usepackage{algorithmic}
\usepackage{graphicx}
\usepackage{textcomp}
\usepackage{xcolor}
\usepackage{url}
\usepackage{multirow}
\usepackage{booktabs}
\usepackage{makecell}
\usepackage{threeparttable}
\usepackage{enumitem}
\usepackage{pifont}

\def\BibTeX{{\rm B\kern-.05em{\sc i\kern-.025em b}\kern-.08em
    T\kern-.1667em\lower.7ex\hbox{E}\kern-.125emX}}

\usepackage{fancyhdr}
\fancypagestyle{firstpage}
{
    \fancyhead[L]{\footnotesize \textcopyright 2026 IEEE.  Personal use of this material is permitted. Permission from IEEE must be obtained for all other uses, in any current or future media, including reprinting/republishing this material for advertising or promotional purposes, creating new collective works, for resale or redistribution to servers or lists, or reuse of any copyrighted component of this work in other works. The paper is accepted at IEEE ATS'26.}
    \fancyhead[R]{}
}

\begin{document}

\title{Carry-Through Checksum: A Lightweight Fault-Detection for CNN Inference at the Edge\thanks{This paper is supported in part by  EU Grant Project 101160182 “TAICHIP”, and the EU Grant 101194287 “NexTArc”.}\vspace{-3mm}
}










\author{\IEEEauthorblockN{Kyrylo Nazarevych\IEEEauthorrefmark{1},
Mohammad Hasan Ahmadilivani\IEEEauthorrefmark{1},
Krister Kaldre\IEEEauthorrefmark{1},
Davide Bertozzi\IEEEauthorrefmark{2},
and Jaan Raik\IEEEauthorrefmark{1}}
\IEEEauthorblockA{\IEEEauthorrefmark{1}\textit{Tallinn University of Technology}, Tallinn, Estonia \\
\{kyrylo.nazarevych, mohammad.ahmadilivani, krister.kaldre, jaan.raik\}@taltech.ee}
\IEEEauthorblockA{\IEEEauthorrefmark{2}\textit{University of Manchester}, Manchester, United Kingdom \\
davide.bertozzi@manchester.ac.uk}\vspace*{-10mm}}

\maketitle
\thispagestyle{firstpage}

\begin{abstract}

Convolutional Neural Networks (CNNs) are increasingly deployed in safety-critical edge applications, where soft errors can silently corrupt inference outputs and lead to unsafe decisions. Such applications typically rely on resource-constrained embedded GPUs, requiring fault detection and mitigation techniques that add minimal compute, memory, and latency overhead while integrating seamlessly with the standard GPU inference pipeline. Existing algorithm-based fault tolerance techniques rely on matrix augmentation and per-operation checksum verification, imposing substantial overhead that is prohibitive for CNN inference on embedded GPUs.

In this work, we propose \textit{carry-through checksum}, a fundamentally new scheme for soft-error detection in CNN inference on embedded GPUs. The method embeds dedicated \textit{carry-through filters} into the convolutional layers, which compute a checksum from the CNN's own operations and propagate it through inference, enabling end-to-end error detection with a single output verification. Experimental results on multiple CNN architectures show that the proposed method detects $95.86\%$ and $86.56\%$ of critical faults for FP32 and FP16, respectively, at almost no additional per-image overhead. Detected faults are mitigated through re-execution, incurring only $2.27\%$ run-time overhead across the entire test set on an NVIDIA Jetson Orin NX GPU. \vspace{-1mm}

\end{abstract}


\section{Introduction} \label{sec:intro}

Convolutional Neural Networks (CNNs) have become the dominant solution for visual perception tasks such as image classification, object detection, and semantic segmentation \cite{younesi2024comprehensive}. These capabilities are increasingly deployed at the edge, embedded directly within safety-critical systems, including autonomous vehicles, unmanned aerial vehicles, and space platforms \cite{gill2025edge}. In such settings, the output of a CNN can directly influence control decisions, so the integrity of every inference is a functional-safety concern: a silently corrupted prediction by soft errors may lead to an unsafe action \cite{ahmadilivani2024systematic, rech2024artificial}. 

At the same time, the hardware executing these models is increasingly susceptible to soft errors; transient faults induced by high-energy particle strikes \cite{hill2021cmos}. A single bit flip in a memory cell, or a computational element, can propagate through the CNN and manifest as Silent Data Corruption (SDC), altering the final prediction without raising any architectural exception \cite{ahmadilivani2026reliability}. Because edge platforms are exposed to harsher operating environments and are often pushed to their energy limits, the probability and impact of such faults are non-negligible. Detecting these errors at inference time, with minimal cost, is therefore essential for ensuring the reliability of CNNs in safety-critical edge deployments \cite{cherezova2026can}.

Among edge compute devices, embedded GPUs, especially the NVIDIA Jetson family, are widely employed in edge safety-critical applications for vision workloads and CNN deployment \cite{kounte2022design}. While they have shown to be susceptible to soft errors \cite{badia2025reliability,veronesi2024cross,he2020fidelity}, embedded GPUs are well-suited to real-time CNN inference under tight power and thermal budgets, owing to their massively parallel architecture, mature software toolchain, and low performance-per-watt. However, the same characteristics constrain any reliability solution: edge GPUs operate under tight memory, energy, and latency constraints. A practical protection scheme for such systems must therefore add limited compute, memory, and latency overhead, while integrating properly with the standard GPU inference pipeline.

To address soft errors in CNNs, Algorithm-Based Fault Tolerance (ABFT) has become a lightweight solution for GPU-accelerated CNNs. Classical ABFT augments matrix operations with redundant checksum rows and columns and verifies them after the computation, through checksum recomputation during inference. Based on this principle, several papers have proposed ABFT mechanisms tailored to CNNs \cite{xu2019safety,ozen2019sanity,zhao2020ft,xue2023approxabft,condia2026ft}, incorporating optimizations and convolution-specific methods to improve fault coverage and resilience. Nevertheless, the need to compute the checksum at the inputs and outputs of each convolution (CONV) creates a mismatch between computation and available resources, introducing additional computational overhead. Also, per-operation fault detection breaks the standard GPU execution flow, prohibiting such techniques in embedded GPUs. To the best of our knowledge, there is no prior work tackling this problem. 

To address this gap, we propose a fundamentally novel frame for error detection in CNN inference for embedded GPUs. For the first time, we propose \textit{carry-through checksum}, a new, lightweight fault detection scheme, aiming to deliver a low-overhead fault detection scheme for CNNs deployed on embedded GPUs. In this method, CNNs' inherent operations compute and propagate a checksum through inference, enabling end-to-end soft error detection with a one-shot verification. The proposed approach replaces the least important channel in each layer by dedicated \textit{carry-through filters} that can inherently calculate the checksum through normal operations of CONV layers, and propagate them through the CNN inference to the output. At the end of an inference, the accumulated checksum is verified by a simple comparison against an expected threshold, and possible errors are detected. 
The main contributions of the paper are as follows:

\begin{itemize}

    \item Introducing \textit{Carry-through checksum}, a novel, lightweight fault detection scheme for CNNs deployed on resource-constrained embedded GPUs, in which the CNNs are modified so that their architecture remains unchanged while computing the checksums by their own operations and propagating them to the outputs, enabling single-shot error detection,
    
    \item Proposing dedicated \textit{carry-through filters} that embed the checksum computation into the CONV layers with zero overhead, as well as proposing two complementary threshold-selection methods (FI-based and a lightweight distribution-based) for identifying an effective detection threshold,
 
    \item Applying the proposed scheme to several CNN architectures with different datasets and datatypes, and deploying them on NVIDIA Jetson Orin NX embedded GPU using TensorRT, and extensively evaluating detection coverage and overhead under fault injection. 
    
\end{itemize}

The evaluation results indicate that the proposed mechanism detects $95.86\%$ and $86.56\%$ of critical faults for FP32 and FP16, respectively, with minimal execution time overhead. Detected faults are mitigated through re-execution, incurring only $2.27\%$ run-time overhead across the entire test set for multiple CNNs on an NVIDIA Jetson Orin NX GPU. 
\section{Carry-Through Checksum} \label{sec:method}

This section introduces the concept of \textit{carry-through checksum} and how to apply it to CNNs. The key idea behind it is to accumulate the Input Feature Maps (IFMaps) of CONV layers through a CNN and propagate them to the final output, without additional operation overhead, and compare them to an expected threshold for fault detection. 

\subsection{Carry-Through Checksum Scheme}

\begin{figure}[t!]
    \centering
    \vspace{-4mm}
    \includegraphics[width=0.35\textwidth]{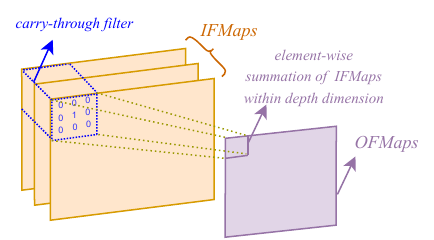}
    \caption{Carry-through checksum in CONV layers. }
    \label{fig:dw-sum}
    \vspace{-7mm}
\end{figure}

The concept of the proposed carry-through checksum for CNNs is to compute, within each CONV layer, the element-wise summation of that layer's IFMaps and accumulate it across the CNN until the final output. Consider a regular CONV layer with $kernel \; size = 3$, $padding=1$ and $stride=1$. To obtain the IFMap summation, we introduce a dedicated weight filter, the \textit{carry-through filter}, whose $3\times3$ kernel is zero everywhere except for a single $1$ at its center, repeated across all input channels, as shown in Fig. \ref{fig:dw-sum}. 

Because a CONV layer inherently sums its products over all input channels, this filter produces, at each spatial position of the Output Feature Map (OFMap), the sum of the co-located IFMap values across the channel (depth) dimension. In other words, the resulting OFMap is a single-channel map whose every element is the element-wise sum of all IFMaps at that position, as illustrated in Fig.~\ref{fig:dw-sum}. Placing the $1$ at the center, rather than using an all-ones kernel, makes this summation non-overlapping. With $padding=1$ and $stride=1$, the center value aligns each output position with exactly one input spatial location, so as the filter slides, every input element contributes to exactly one output element. An all-ones kernel, by contrast, would cause neighbouring windows to overlap and count the same input elements multiple times. The \textit{carry-through filter} therefore yields a clean, one-to-one summation across the channel dimension, with no double counting.

\begin{figure}[t!]
    \centering
    \vspace{-2mm}
    \includegraphics[width=0.5\textwidth]{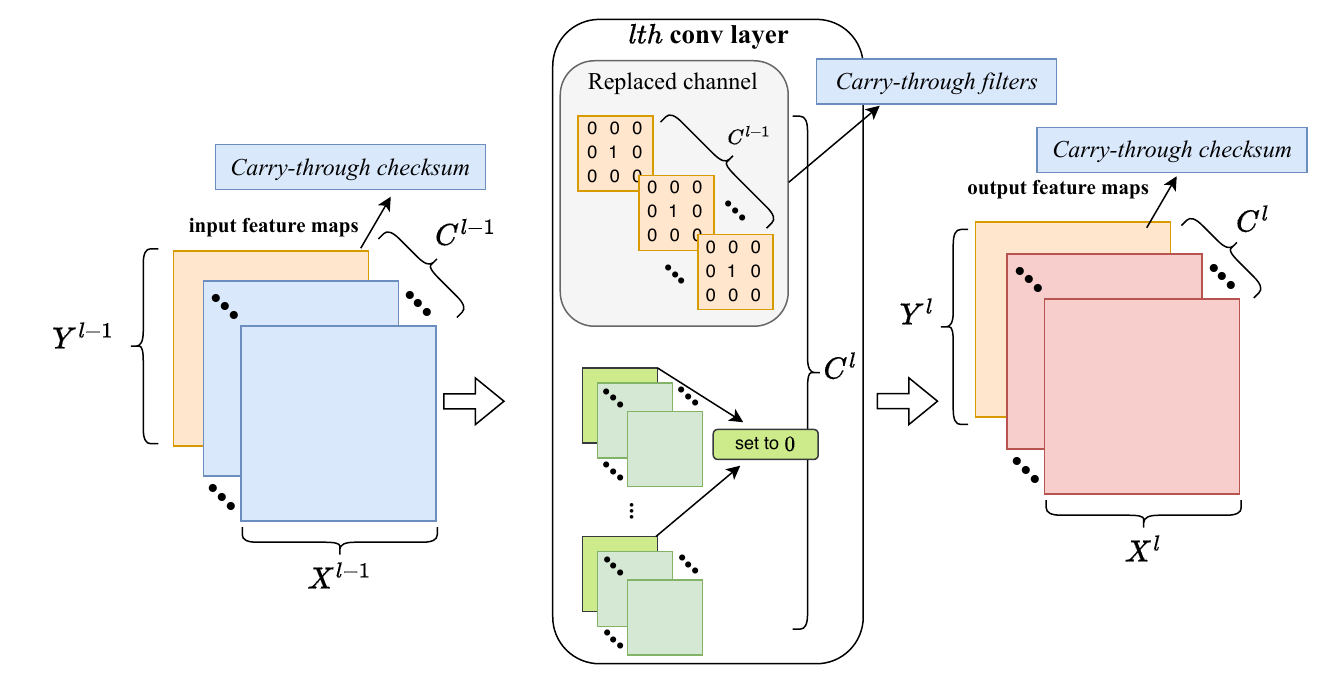}
    \caption{Carry-through checksum in a CNN. }
    \label{fig:carry-through}
    \vspace{-2mm}
\end{figure}

To embed \textit{carry-through filters} into a CNN, we modify each CONV layer at two levels, as shown in Fig. \ref{fig:carry-through}: 

\textbf{1. Output-channel level:} We structurally prune the least important 3D output-channel filter, and append the \textit{carry-through filter}, as the new last output channel. This added channel produces the \textit{carry-through checksum} in the layer's OFMaps. Note that in the structured pruning, the corresponding 2D filters in the input channel in the next layer is also pruned. 
To identify the least important output channel to prune, we compute, for each channel, the combined \textit{L1-norm} of its associated filters in the current layer's output and the next layer's input, and prune the channel with the smallest value.

\textbf{2. Input-channel level:} Because the IFMaps now contain the checksum channel propagated from the previous layer, this incoming checksum must not leak into the layer's normal OFMaps. The previous layer appends its checksum as the \emph{last} output channel, so it arrives as the \emph{last} input channel in the current layer. Accordingly, in every normal output-channel filter, we set the 2D kernel slice that multiplies this last (checksum) input channel to all zeros. Since one 2D filter in each input channel has already been pruned, the new filter does not change the original size of the layer.
As a result, the incoming checksum passes through the layer without contributing to the normal OFMap computation.

Since only the output channel and the corresponding 2D slice per filter are pruned, and each is replaced by a dedicated new filter or slice, the number of output channels and the per-filter dimensions are preserved. The layer, therefore, incurs no additional compute or memory overhead in CONV layers. As shown in Fig.~\ref{fig:carry-through}, the \textit{carry-through filter} produces one additional output channel that holds the \textit{carry-through checksum}, and the depth of each \textit{carry-through filter} equals the number of input channels to the CONV layer, so that it spans all of them. The same scheme is applied to Fully-Connected (FC) layers as well, while we prune one neuron with the least \textit{L1-norm} result and add one neuron whose weights are $1$. After CNN modification, if the test accuracy drops more than $1\%$ by pruning, we conduct a lightweight fine-tuning to retain it.

This scheme is also extended to non-regular CONV layers. In ResNet-like architectures, where residual channels are present, the \textit{carry-through filters} are applied to all CONV layers except the layers with an input residual connection. In other words, the first layer in each residual block does not contain the \textit{carry-through checksum}, yet its least important channel is pruned, and the last spatial channel contains only $0$. In this way, the incoming IFMaps will not be accumulated twice. In MobileNet-like architectures, where \textit{ReLU6} is present, the accumulated checksums may get saturated, since they exceed $6$. Therefore, in such CNNs, we replace them with normal \textit{ReLU} in the CNN to maintain the efficacy of the method. 

In CNNs, the Batch Normalization (BN) function is also present, and performs Eq. \eqref{eq:bn}. In this equation, $x$ is an input batch, $E[.]$ is their mean, $Var[.]$ is their variance, $\gamma$ and $\beta$ are learning parameters. In our method, to ensure that the BN function does not modify the \textit{carry-through checksum}, we set the last values in the BN layers as follows: $E(x) = 0$, $Var[x] = 1$, $\gamma = 1$, and $\beta = 0$. This ensures that the BN layer does not modify the values of those accumulated checksum channels, essentially acting as an identity mapping on the \textit{carry-through checksum}. 

\begin{equation}
    BN(x) = \frac{x - E[x]}{\sqrt{Var[x] + \epsilon}} \times \gamma + \beta
    \label{eq:bn}
\end{equation}

\subsection{Fault Detection and Mitigation}

To perform fault detection at inference time, we augment the final Fully-Connected (FC) layer with a single additional neuron, called \textit{accumulated-checksum neuron}. Every input to the FC layer is connected to this neuron with a fixed weight of $1$, so that the neuron computes the sum of all incoming activations, including the \textit{carry-through checksum} values propagated through the CNN. The \textit{accumulated-checksum neuron} thus aggregates the \textit{carry-through checksum} at the CNN output, where it serves as the indicator for error detection.

In a fault-free inference, this neuron takes an expected reference value determined by the propagated checksum. If a soft error occurs during inference, its effect is carried forward by the \textit{carry-through checksum} and is reflected as a deviation in the \textit{accumulated-checksum neuron}. We therefore define a detection threshold $\tau$; if the deviation of the \textit{accumulated-checksum neuron}'s value from its expected reference exceeds $\tau$, an error-detection flag is raised, and the inference is flagged as corrupted. The objective of the error detection is to \textbf{identify the critical faults}, i.e., the faults that lead to a change of the golden classification. Upon detection, the inference is re-executed to recover a correct output. In the re-execution, we ignore the detection result, as it is unlikely to have faults in two consecutive inferences.

\subsection{Detection Threshold Optimization} \label{subsec:threshold}

To determine the detection threshold $\tau$, we propose two complementary methods: (1) FI-based and (2) distribution-based. In the FI-based method, we run FI campaigns using the training set and assign a $\tau$, while False Positive Rate (FPR) is less than $\alpha \%$; the fraction of non-critical faults (those not changing the golden classification) that are classified wrongly as critical by \textit{accumulated-checksum neuron} exceeding $\tau$. We explore different values for $\alpha$ to find its best value. 

In the distribution-based method, we characterize the statistical distribution of the \textit{accumulated-checksum neuron}'s value over fault-free inferences on the training set, and assign a $\tau$ where less than $\alpha\%$ of the fault-free executions are wrongly flagged as critical. This would minimize the false alarms (FPR) in the execution. This method requires no FI and depends only on the fault-free behavior of the CNN equipped with \textit{carry-through filters}. We explore $\alpha$ to find its best value. 

\section{Experiments} \label{sec:results}

\subsection{Experimental Setup}

In this work, we consider soft errors in the parameters of the CNNs. Soft errors can cause bit-flips in the parameters that are stored in the memory of embedded GPUs and change the values of parameters. We perform Fault Injection (FI) campaigns on the parameters of the CNNs. We inject numerous random single bit-flips with the test set until the average results reach a $95\%$ confidence level with $1\%$ error margin. Single faults are injected during inference in PyTorch on test set.
To evaluate the efficacy of the proposed method under FI, we measure two metrics:

\begin{itemize}

\item \textit{True-Positive Ratio (TPR)}: the fraction of critical faults (those that alter the golden classification) that are correctly detected;

\item \textit{False Positive Ratio (FPR)}: the fraction of non-critical faults (those that do not change the golden classification) that are wrongly detected as critical.
\end{itemize}

TPR and FPR are explored with various $\alpha$s, including $\{0.1\%, 1\%, 5\%\}$, for both FI-based and distribution-based threshold optimization methods described in Subsection \ref{subsec:threshold}. To identify the best $\alpha$, we derive Youden's J statistic. It is obtained by $TPR - FPR$, representing how accurately the method captures the true positive critical cases. 

We perform the experiments on multiple CNN architectures: VGG-11, VGG-19, ResNet-20, ResNet-56, MobileNet-V2-x0.5 and MobileNet-V2-x1.4, pre-trained\footnote{Loaded from \url{https://github.com/chenyaofo/pytorch-cifar-models}} on two datasets: CIFAR-10 and CIFAR-100. We consider Floating Point 32 (FP32) and FP16 datatypes in the experiments. After applying the \textit{carry-through filters} to the CNNs, we conduct a lightweight fine-tuning to retain the accuracy. Fine-tuning is performed with $25$ epochs, and a learning rate $0.001$, with the SGD optimizer.

To evaluate the proposed mechanism, we deploy the CNNs on an NVIDIA Jetson Orin NX (16 GB) GPU using TensorRT 8.5.2. TensorRT is used to generate inference engine files in both FP32 and FP16 precisions, which are then executed on the GPU via a CUDA 11.4 and C++ implementation. We assess the overhead of \textit{carry-through checksum} modification on CNNs, and based on that we project the end-to-end performance of the complete fault detection and mitigation pipeline, in which a detected fault triggers re-execution of the inference. We report the resulting overhead over a full pass on the test set, assessed both with and without FI.

\subsection{Experimental Results}

\subsubsection{Impact of Carry-Through Filters on Baseline accuracy}

Because the proposed scheme modifies the filters of the CNN, it may incur a reduction in accuracy. For models whose accuracy drops by more than $1\%$ after pruning, we apply fine-tuning to recover their accuracy. Table~\ref{tab:accuracy-impact} reports the Top-1 test accuracy of the CNNs equipped with \textit{carry-through filters} in FP32, both after pruning (without fine-tuning) and after fine-tuning. As shown, pruning has a negligible effect on the VGG variants, which therefore require no fine-tuning. In contrast, the ResNet and MobileNet variants are more sensitive to channel pruning and do require fine-tuning. In all cases, the accuracy of the final carry-through-protected models remains within $1\%$ of the corresponding baselines. It is worth mentioning that the accuracy of CNNs with FP16 datatype is around $0.1\%$ different than FP32 results.

\begin{table}[b!]
\centering
\vspace{-7mm}
\caption{Impact of carry-through filters on the CNNs' accuracy.}
\label{tab:accuracy-impact}
\begin{threeparttable}
\resizebox{0.48\textwidth}{!}{
\begin{tabular}{llccc}
\toprule
Dataset & CNN & \makecell{Baseline\\accuracy} &  \makecell{Pruned \\ accuracy} & \makecell{Fine-tuned \\ accuracy}   \\
\midrule
\multirow{6}{*}{\makecell{CIFAR\\-10}}
 & VGG-11             & 92.79\% & 92.61\%  & 92.61\%\tnote{*}  \\
 & VGG-19             & 93.91\% & 93.73\%  & 93.73\%\tnote{*} \\
 & ResNet-20          & 92.59\% & 85.06\%  & 91.71\%  \\
 & ResNet-56          & 94.38\% & 92.02\%  & 93.53\%  \\
 & MobileNet-V2-x0.5 & 93.12\%  & 90.55\%  & 92.24\%  \\
 & MobileNet-V2-x1.4 & 94.21\%  & 94.07\%  & 94.07\%\tnote{*}  \\
\midrule
\multirow{6}{*}{\makecell{CIFAR\\-100}}
 & VGG-11             & 70.79\% & 70.69\%  & 70.69\%\tnote{*}  \\
 & VGG-19             & 73.85\% & 73.43\%  & 73.43\%\tnote{*} \\
 & ResNet-20          & 68.83\% & 38.11\%  & 67.88\%  \\
 & ResNet-56          & 72.60\% & 18.61\%  & 71.65\%  \\
 & MobileNet-V2-x0.5 & 71.13\% &  68.92\%  & 70.38\%  \\
 & MobileNet-V2-x1.4 & 76.34\% &  75.23\%  & 75.38\%  \\
\bottomrule
\end{tabular}
}
\begin{tablenotes}
\footnotesize
\item[*] No fine-tuning applied.
\end{tablenotes}
\end{threeparttable}
\end{table}

\subsubsection{Detection Ratio of Carry-Through Checksum}

\begin{table*}[t]
\centering
\caption{TPR and FPR of \textit{carry-through checksum} for FP32 datatype, under the FI experiments with the exploration of $\alpha$. }
\label{tab:detection-ratio-fp32}
\resizebox{\textwidth}{!}{%
\begin{tabular}{llcccccccccccc}
\toprule
 & & \multicolumn{6}{c}{FI-based} & \multicolumn{6}{c}{Distribution-based} \\
\cmidrule(lr){3-8} \cmidrule(lr){9-14}
 & & \multicolumn{3}{c}{TPR} & \multicolumn{3}{c}{FPR}
   & \multicolumn{3}{c}{TPR} & \multicolumn{3}{c}{FPR} \\
\cmidrule(lr){3-5} \cmidrule(lr){6-8} \cmidrule(lr){9-11} \cmidrule(lr){12-14}
Dataset & CNN & $\alpha=0.1\%$ & $\alpha=1\%$ & $\alpha=5\%$ & $\alpha=0.1\%$ & $\alpha=1\%$ & $\alpha=5\%$ & $\alpha=0.1\%$ & $\alpha=1\%$ & $\alpha=5\%$ & $\alpha=0.1\%$ & $\alpha=1\%$ & $\alpha=5\%$ \\
\midrule
\multirow{6}{*}{CIFAR-10}
 & VGG-11            & 59.65\% & 98.28\% & \textbf{98.33\%} & \textbf{0.10\%} & 1.35\% & 4.74\% & 98.25\% & 98.28\% & \textbf{98.34\%} & \textbf{0.25\%} & 1.35\% & 5.10\% \\
 & VGG-19            & 37.96\% & 99.28\% & \textbf{99.32\%} & \textbf{0.10\%} & 0.83\% & 4.77\% & 99.27\% & 99.29\% & \textbf{99.32\%} & \textbf{0.34\%} & 0.95\% & 4.77\% \\
 & ResNet-20         & 96.54\% & 96.60\% & \textbf{96.70\%} & \textbf{0.20\%} & 1.04\% & 5.14\% & 96.55\% & 96.60\% & \textbf{96.69\%} & \textbf{0.22\%} & 1.07\% & 5.02\% \\
 & ResNet-56         & 98.00\% & 98.06\% & \textbf{98.11\%} & \textbf{0.18\%} & 1.18\% & 5.10\% & 98.03\% & 98.06\% & \textbf{98.11\%} & \textbf{0.24\%} & 1.10\% & 4.74\% \\
 & MobileNet-V2-x0.5 & 41.48\% & 98.10\% & \textbf{98.17\%} & \textbf{0.11\%} & 1.14\% & 4.52\% & 98.09\% & 98.11\% & \textbf{98.17\%} & \textbf{0.35\%} & 1.47\% & 4.71\% \\
 & MobileNet-V2-x1.4 & 40.59\% & 99.70\% & \textbf{99.70\%} & \textbf{0.10\%} & 0.83\% & 3.49\% & 99.69\% & 99.70\% & \textbf{99.71\%} & \textbf{0.33\%} & 1.03\% & 4.08\% \\
\midrule
\multirow{6}{*}{CIFAR-100}
 & VGG-11            & 96.36\% & 96.38\% & \textbf{96.48\%} & \textbf{0.13\%} & 1.02\% & 4.55\% & 96.36\% & 96.38\% & \textbf{96.49\%} & \textbf{0.13\%} & 1.15\% & 4.66\% \\
 & VGG-19            & 98.54\% & 98.55\% & \textbf{98.59\%} & \textbf{0.07\%} & 0.84\% & 4.28\% & 98.54\% & 98.55\% & \textbf{98.58\%} & \textbf{0.07\%} & 0.80\% & 4.15\% \\
 & ResNet-20         & 85.78\% & 85.96\% & \textbf{86.35\%} & \textbf{0.30\%} & 1.29\% & 5.55\% & 85.77\% & 85.96\% & \textbf{86.35\%} & \textbf{0.19\%} & 1.22\% & 5.54\% \\
 & ResNet-56         & 92.01\% & 92.11\% & \textbf{92.75\%} & \textbf{0.08\%} & 0.64\% & 4.38\% & 92.05\% & 92.12\% & \textbf{92.75\%} & \textbf{0.09\%} & 0.76\% & 4.38\% \\
 & MobileNet-V2-x0.5 & 88.95\% & 89.05\% & \textbf{89.53\%} & \textbf{0.07\%} & 0.97\% & 4.74\% & 88.95\% & 89.06\% & \textbf{89.53\%} & \textbf{0.07\%} & 1.03\% & 4.77\% \\
 & MobileNet-V2-x1.4 & 97.21\% & 97.24\% & \textbf{97.36\%} & \textbf{0.03\%} & 0.71\% & 4.32\% & 97.21\% & 97.24\% & \textbf{97.36\%} & \textbf{0.03\%} & 0.68\% & 4.32\% \\
 \midrule
 \multirow{2}{*}{} 
 & Average    & 77.76\% & 95.78\% & \textbf{95.95\%} & \textbf{0.12\%} & 0.99\% & 4.63\% & 95.78\% & 95.86\% & \textbf{95.95\%} & \textbf{0.19\%} & 1.05\% & 4.69\% \\
 \midrule
 & Youden's J & \multicolumn{6}{c}{$\alpha=0.1\%$: 77.64\% \quad $\alpha=1\%$: \textbf{94.79\%} \quad $\alpha=5\%$: 91.32\%} & \multicolumn{6}{c}{$\alpha=0.1\%$: \textbf{95.59\%} \quad $\alpha=1\%$: 94.81\% \quad $\alpha=5\%$: 91.26\%} \\
\bottomrule
\end{tabular}%
}
\vspace{-3mm}
\end{table*}

\begin{table*}[t]
\centering
\caption{TPR and FPR of \textit{carry-through checksum} for FP16 datatype, under the FI experiments with the exploration of $\alpha$.}
\label{tab:detection-ratio-fp16}
\resizebox{\textwidth}{!}{%
\begin{tabular}{llcccccccccccc}
\toprule
 & & \multicolumn{6}{c}{FI-based} & \multicolumn{6}{c}{Distribution-based} \\
\cmidrule(lr){3-8} \cmidrule(lr){9-14}
 & & \multicolumn{3}{c}{TPR} & \multicolumn{3}{c}{FPR}
   & \multicolumn{3}{c}{TPR} & \multicolumn{3}{c}{FPR} \\
\cmidrule(lr){3-5} \cmidrule(lr){6-8} \cmidrule(lr){9-11} \cmidrule(lr){12-14}
Dataset & CNN & $\alpha=0.1\%$ & $\alpha=1\%$ & $\alpha=5\%$ & $\alpha=0.1\%$ & $\alpha=1\%$ & $\alpha=5\%$ & $\alpha=0.1\%$ & $\alpha=1\%$ & $\alpha=5\%$ & $\alpha=0.1\%$ & $\alpha=1\%$ & $\alpha=5\%$ \\
\midrule
\multirow{6}{*}{CIFAR-10}
 & VGG-11            & 61.39\% & 89.98\% & \textbf{91.63\%} & \textbf{0.09\%} & 1.32\% & 4.88\% & 88.24\% & 90.19\% & \textbf{91.72\%} & \textbf{0.38\%} & 1.57\% & 5.37\% \\
 & VGG-19            & 70.60\% & 76.91\% & \textbf{79.82\%} & \textbf{0.10\%} & 0.81\% & 4.94\% & 73.48\% & 77.11\% & \textbf{79.82\%} & \textbf{0.23\%} & 0.92\% & 4.94\% \\
 & ResNet-20         & 38.73\% & 95.53\% & \textbf{95.94\%} & \textbf{0.12\%} & 0.95\% & 4.88\% & 95.15\% & 95.59\% & \textbf{95.94\%} & \textbf{0.38\%} & 1.21\% & 4.88\% \\
 & ResNet-56         & 37.21\% & 96.00\% & \textbf{96.49\%} & \textbf{0.10\%} & 1.00\% & 4.92\% & 95.50\% & 96.05\% & \textbf{96.48\%} & \textbf{0.41\%} & 1.26\% & 4.82\% \\
 & MobileNet-V2-x0.5 & 50.14\% & 91.82\% & \textbf{93.80\%} & \textbf{0.13\%} & 1.11\% & 4.37\% & 90.34\% & 92.46\% & \textbf{93.93\%} & \textbf{0.50\%} & 1.67\% & 4.98\% \\
 & MobileNet-V2-x1.4 & 59.54\% & 89.66\% & \textbf{91.84\%} & \textbf{0.11\%} & 0.82\% & 3.57\% & 87.82\% & 90.16\% & \textbf{92.09\%} & \textbf{0.39\%} & 1.16\% & 4.27\% \\
\midrule
\multirow{6}{*}{CIFAR-100}
 & VGG-11            & 86.76\% & 88.02\% & \textbf{89.33\%} & \textbf{0.15\%} & 1.04\% & 4.57\% & 86.88\% & 88.14\% & \textbf{89.38\%} & \textbf{0.15\%} & 1.22\% & 4.70\% \\
 & VGG-19            & 69.90\% & 75.62\% & \textbf{77.93\%} & \textbf{0.08\%} & 0.84\% & 4.34\% & 73.10\% & 75.90\% & \textbf{77.93\%} & \textbf{0.18\%} & 1.08\% & 4.34\% \\
 & ResNet-20         & 85.47\% & 85.93\% & \textbf{86.48\%} & \textbf{0.23\%} & 1.29\% & 5.48\% & 85.47\% & 85.92\% & \textbf{86.48\%} & \textbf{0.23\%} & 1.25\% & 5.48\% \\
 & ResNet-56         & 90.29\% & 90.98\% & \textbf{92.06\%} & \textbf{0.09\%} & 0.66\% & 4.37\% & 90.45\% & 91.04\% & \textbf{92.04\%} & \textbf{0.10\%} & 0.76\% & 4.26\% \\
 & MobileNet-V2-x0.5 & 79.17\% & 81.07\% & \textbf{83.25\%} & \textbf{0.08\%} & 0.90\% & 4.70\% & 79.27\% & 81.19\% & \textbf{83.28\%} & \textbf{0.10\%} & 1.01\% & 4.77\% \\
 & MobileNet-V2-x1.4 & 71.79\% & 74.59\% & \textbf{78.26\%} & \textbf{0.08\%} & 0.56\% & 4.16\% & 71.23\% & 75.02\% & \textbf{78.31\%} & \textbf{0.06\%} & 0.67\% & 4.26\% \\
 \midrule
 \multirow{2}{*}{} 
 & Average    & 66.75\% & 86.34\% & \textbf{88.07\%} & \textbf{0.11\%} & 0.94\% & 4.60\% & 84.74\% & 86.56\% & \textbf{88.12\%} & \textbf{0.26\%} & 1.15\% & 4.76\% \\
 \midrule
 & Youden's J & \multicolumn{6}{c}{$\alpha=0.1\%$: 66.64\% \quad $\alpha=1\%$: \textbf{85.40\%} \quad $\alpha=5\%$: 83.47\%} & \multicolumn{6}{c}{$\alpha=0.1\%$: 84.48\% \quad $\alpha=1\%$: \textbf{85.41\%} \quad $\alpha=5\%$: 83.36\%} \\
\bottomrule
\end{tabular}%
}
\vspace{-3mm}
\end{table*}

This subsection presents the detection ratio results for the proposed method while exploring threshold optimization methods. Table \ref{tab:detection-ratio-fp32} and Table \ref{tab:detection-ratio-fp16} present the obtained detection ratios (TPR and FPR) for FP32 and FP16 datatypes, respectively, under both FI-based and distribution-based thresholds optimizations. The following observations can be highlighted:


\ding{202} \textbf{Threshold exploration:} To achieve effective fault detection in \textit{carry-through checksum}, we investigate three candidate values of $\alpha$ and evaluate their impact on TPR and FPR under FI. An effective threshold must simultaneously maximize TPR to ensure reliable detection of critical faults and minimize FPR to limit false alarms and performance overhead. Our results indicate that $\alpha=5\%$ consistently yields the highest TPR, albeit at the cost of a substantially elevated FPR, whereas $\alpha=0.1\%$ achieves the lowest FPR but suffers from a markedly reduced TPR. Since these two objectives are inherently competing, we adopt Youden's J statistic as a unified metric to identify the best trade-off. This metric identifies $\alpha=1\%$ as the most effective operating point, as it consistently maximizes the balance between sensitivity and false-alarm rate across all configurations. This trend holds across both the FI-based and distribution-based detection methods, as well as across both the FP32 and FP16 datatypes, wherein $\alpha=1\%$ maintains a favorable combination of high TPR and low FPR.

\ding{203} \textbf{Threshold optimization method}: 
At $\alpha=1\%$, both the FI-based and distribution-based methods demonstrate comparably effective detection performance, with the observed differences between the two remaining marginal in most configurations. Nevertheless, the FI-based method entails considerable computational complexity, as it requires extensive FI campaigns to empirically optimize the threshold, resulting in substantially longer execution times. In contrast, the distribution-based technique derives the threshold analytically from the underlying data distribution, offering a simple and computationally efficient alternative that circumvents the need for FI experiments. Given this favorable trade-off between detection performance and computational cost, we conclude that \textbf{the \textit{carry-through checksum} employing distribution-based threshold optimization with $\alpha=1\%$ constitutes the practical and effective technique proposed in this paper.}

\ding{204} \textbf{Impact of \textit{carry-through checksum}:} When employing the \textit{carry-through checksum} with a threshold derived via the distribution-based method at $\alpha=1\%$, the evaluated CNNs exhibit a remarkable detection capability across a substantial portion of critical faults. Across the full set of experiments spanning multiple CNN architectures and datasets, this configuration yields, for the FP32 datatype, a TPR ranging from $85.96\%$ to $99.7\%$ (average $95.86\%$), indicating that the vast majority of critical faults are correctly identified, while the corresponding FPR remains confined to the range of $0.68\%$ to $1.35\%$ (average $1.05\%$), reflecting a comparably low rate of erroneous flagging among non-critical faults. For the FP16 datatype, the TPR ranges from $75.02\%$ to $96.05\%$ (average $86.56\%$), while the FPR ranges $0.67\%$ to $1.67\%$ (average $1.15\%$). Overall, results underscore the effectiveness of the proposed method in achieving consistently high detection rates of critical faults across CNN architectures, while maintaining a low incidence of false alarms.

\ding{205} \textbf{Impact of CNNs depth, datasets, and datatype:} Comparing the same CNN architectures across different datasets reveals that detection ratios are marginally lower for CIFAR-100 relative to CIFAR-10. This can be attributed to the larger number of output classes in CIFAR-100, which forces the same architecture to spread its activation value distributions over a wider range, consequently yielding a noticeably higher threshold value for a given $\alpha$, thereby masking a larger fraction of critical faults. Nonetheless, the detection ratio remains consistently high, indicating that this effect does not substantially compromise the overall effectiveness of the method. Comparing FP32 and FP16 precisions, the average TPR under FP16 is $9.3\%$ lower than that under FP32. This degradation stems from the smaller range of representable values in FP16 and the rounding error for checksum calculation. Therefore, errors are more likely to be less than the detection threshold and escape detection while still being able to alter the classification. Despite this reduction, the detection performance under FP16 remains substantial, with critical faults correctly identified in up to $96.05\%$ of cases. Furthermore, a cross-comparison among CNN architectures indicates that the detection ratio remains consistently high across both shallow and deep CNN variants, without exhibiting any noticeable dependence on network depth or parameter count. This suggests that the architectural depth and size of the CNN do not influence the efficacy of the \textit{carry-through checksum}. Collectively, these observations demonstrate that \textbf{the proposed method is highly effective, generalizable, and scalable across diverse datasets, numerical datatypes, and CNN architectures}.


\begin{figure*}[h!]
\centering
\includegraphics[width=\textwidth]{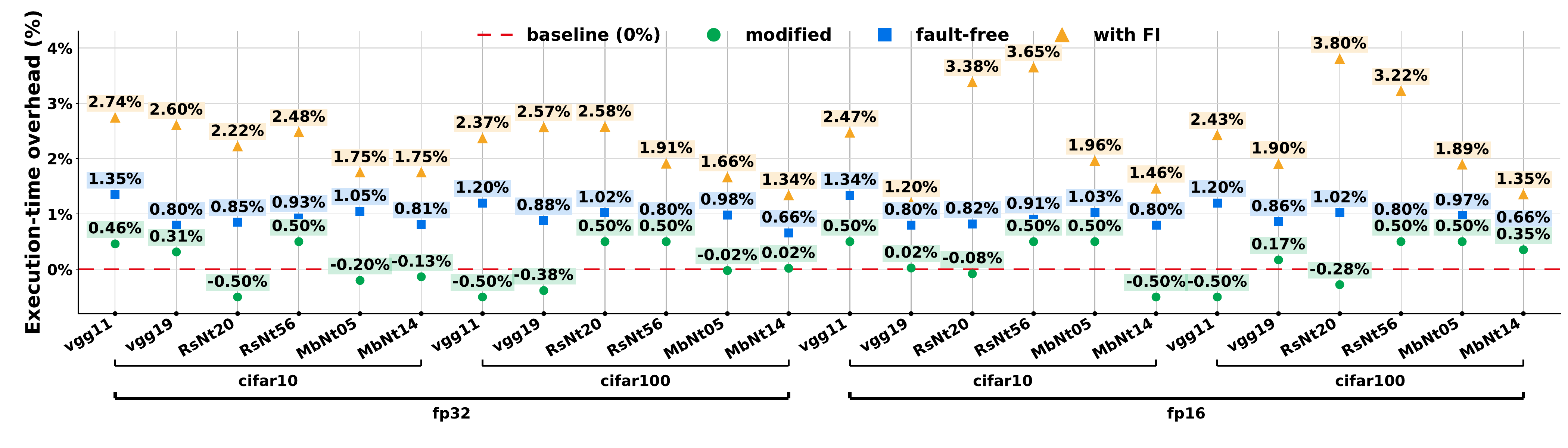}
\vspace{-8mm}
\caption{Execution time overhead of all CNNs with FP32 and FP16 precisions, compared to the baseline CNNs.}
\vspace{-5mm}
\label{fig:latency-overhead}
\end{figure*}

\subsubsection{Execution-Time Overhead Analysis}

To quantify the impact of the proposed method on inference latency in resource-constrained devices, we compare the execution time of baseline CNNs against their modified counterparts employing the \textit{carry-through checksum}, evaluated on an NVIDIA Jetson Orin NX embedded GPU. As shown in Fig. \ref{fig:latency-overhead}, the modified CNNs incorporating \textit{carry-through filters} exhibit an execution-time overhead ranging from $-0.5\%$ to $+0.5\%$, with an average below $0.1\%$; a variation that is effectively negligible and largely attributable to inherent run-time fluctuations of the GPU deployment rather than to the method itself. This minimal overhead arises because the proposed approach only augments the final FC layer with a small number of additional weights, leaving the architecture of all preceding layers entirely unaltered; consequently, the resulting inference latency remains nearly indistinguishable from that of the baseline CNN. It is worth mentioning that augmenting the last layer with the additional \textit{accumulated-checksum neuron} adds a negligible memory overhead to a CNN, i.e., $<10^{-3}\%$.

Considering the overhead introduced by fault detection and re-execution-based mitigation, a complete fault-free inference over the entire test set incurs an execution time overhead ranging from $0.66\%$ to $1.35\%$, with an average of $0.93\%$. Given that the intrinsic overhead of the modified CNN is negligible, this cost is attributable mainly to re-executions triggered by false alarms occurring during the fault-free operation.

We further characterize the execution-time overhead under the combined effect of fault detection and mitigation in the presence of injected faults. Under FI campaigns under the test inputs, the system incurs an overhead ranging from $1.2\%$ to $3.8\%$, with an average of $2.27\%$, caused by both critical and non-critical faults in which a detection flag triggers a single re-execution. This result confirms the efficiency of the proposed detection and mitigation scheme, which introduces only a modest overhead relative to the substantial gain in fault-detection capability.

These results demonstrate that the \textit{carry-through checksum} imposes minimal execution-time overhead on embedded GPU platforms while achieving highly effective detection of critical faults, thereby establishing it as a practical and low-cost solution for reliable CNN inference in resource-constrained, safety-critical deployments.

\section{Conclusions} \label{sec:conclusion}

This work presents \textit{carry-through checksum}, a novel scheme for soft error fault detection in CNNs for resource-constrained devices. We introduce dedicated \textit{carry-through filters} without changing the structure and operations of the CNNs, inherently calculating and propagating the checksum of IFMaps within an inference, and detecting critical faults at the output of the CNNs. Furthermore, through an extensive exploration, we demonstrate that the lightweight distribution-based method can identify the detection threshold fast and effectively.  

The evaluation results indicate that the proposed mechanism detects on average $95.86\%$ and $86.56\%$ of critical faults for FP32 and FP16, respectively, with minimal memory and execution time overhead, across several CNNs. Detected faults are mitigated through re-execution, incurring only $2.27\%$ run-time overhead across the entire test set for multiple CNNs on an NVIDIA Jetson Orin NX GPU.

\bibliographystyle{IEEEtran}
\bibliography{ref.bib}

@article{ahmadilivani2024systematic,
  title={A systematic literature review on hardware reliability assessment methods for deep neural networks},
  author={Ahmadilivani, Mohammad Hasan and Taheri, Mahdi and Raik, Jaan and Daneshtalab, Masoud and Jenihhin, Maksim},
  journal={ACM Computing Surveys},
  volume={56},
  number={6},
  pages={1--39},
  year={2024},
  publisher={ACM New York, NY}
}

@article{rech2024artificial,
  title={Artificial neural networks for space and safety-critical applications: Reliability issues and potential solutions},
  author={Rech, Paolo},
  journal={IEEE Transactions on Nuclear Science},
  year={2024},
  publisher={IEEE}
}

@article{younesi2024comprehensive,
  title={A comprehensive survey of convolutions in deep learning: Applications, challenges, and future trends},
  author={Younesi, Abolfazl and Ansari, Mohsen and Fazli, Mohammadamin and Ejlali, Alireza and Shafique, Muhammad and Henkel, J{\"o}rg},
  journal={IEEE Access},
  volume={12},
  pages={41180--41218},
  year={2024},
  publisher={IEEE}
}

@article{cherezova2026can,
  title={Can Model-Level Fault Tolerance be Enough for DNN Hardware Accelerators? A Study},
  author={Cherezova, Natalia and Ahmadilivani, Mohammad Hasan and Raik, Jaan and Daneshtalab, Masoud and Jenihhin, Maksim},
  journal={IEEE Design \& Test},
  year={2026},
  publisher={IEEE}
}

@inproceedings{condia2026ft,
  title={FT-Sparse: Algorithm-Based Fault Tolerance for Sparse CNNs Using Structured Sparsity in GPUs},
  author={Condia, Josie E Rodriguez and Ahmadilivani, Mohammad Hasan and Raik, Jaan and Jenihhin, Maksim and Reorda, Matteo Sonza},
  booktitle={2026 IEEE 44th VLSI Test Symposium (VTS)},
  pages={1--7},
  year={2026},
  organization={IEEE}
}

@article{gill2025edge,
  title={Edge AI: A taxonomy, systematic review and future directions},
  author={Gill, Sukhpal Singh and Golec, Muhammed and Hu, Jianmin and Xu, Minxian and Du, Junhui and Wu, Huaming and Walia, Guneet Kaur and Murugesan, Subramaniam Subramanian and Ali, Babar and Kumar, Mohit and others},
  journal={Cluster Computing},
  volume={28},
  number={1},
  pages={18},
  year={2025},
  publisher={Springer}
}

@article{hill2021cmos,
  title={CMOS reliability from past to future: A survey of requirements, trends, and prediction methods},
  author={Hill, Ian and Chanawala, Parvez and Singh, Rohit and Sheikholeslam, S Arash and Ivanov, Andr{\'e}},
  journal={IEEE Transactions on Device and Materials Reliability},
  volume={22},
  number={1},
  pages={1--18},
  year={2021},
  publisher={IEEE}
}

@inproceedings{ahmadilivani2026reliability,
  title={Reliability Assessment of Deep Neural Networks and Accelerators Across Design Stages},
  author={Ahmadilivani, Mohammad Hasan and Cherezova, Natalia and Gla{\ss}, Michael and Guerrero-Balaguera, Juan-David and Jenihhin, Maksim and Kritikakou, Angeliki and Sierra, Robert Limas and Poelhs, Leticia Bolzani and Raik, Jaan and Roquet, Lucas and others},
  booktitle={2026 IEEE 27th Latin American Test Symposium (LATS)},
  pages={1--10},
  year={2026},
  organization={IEEE}
}

@inproceedings{kounte2022design,
  title={Design and development of autonomous driving car using NvidiaJetson Nano developer kit},
  author={Kounte, Manjunath R and Harshvardhan, Vs and Kumari, Ayushi and Dhruv, S and others},
  booktitle={2022 IEEE 4th International Conference on Cybernetics, Cognition and Machine Learning Applications (ICCCMLA)},
  pages={486--489},
  year={2022},
  organization={IEEE}
}

@article{zhao2020ft,
  title={FT-CNN: Algorithm-based fault tolerance for convolutional neural networks},
  author={Zhao, Kai and Di, Sheng and Li, Sihuan and Liang, Xin and Zhai, Yujia and Chen, Jieyang and Ouyang, Kaiming and Cappello, Franck and Chen, Zizhong},
  journal={IEEE Transactions on Parallel and Distributed Systems},
  volume={32},
  number={7},
  pages={1677--1689},
  year={2020},
  publisher={IEEE}
}

@article{xue2023approxabft,
  title={ApproxABFT: Approximate Algorithm-Based Fault Tolerance for Neural Network Processing},
  author={Xue, Xinghua and Liu, Cheng and Min, Feng and Luo, Tao and Han, Yinhe},
  journal={arXiv preprint arXiv:2302.10469},
  year={2023}
}

@article{badia2025reliability,
  title={Reliability of Vision Transformers and CNNs on Edge AI systems under neutron radiation},
  author={Badia, Jose M and Martin-Salinas, Ignacio and Leon, German and Amor-Martin, Adrian and Frias-Dominguez, Lester and Belloch, Jose A and Garcia-Valderas, Mario and Lindoso, Almudena and Cazzaniga, Carlo and Entrena, Luis},
  journal={IEEE Transactions on Nuclear Science},
  volume={72},
  number={8},
  pages={2706--2716},
  year={2025},
  publisher={IEEE}
}

@inproceedings{xu2019safety,
  title={Safety design of a convolutional neural network accelerator with error localization and correction},
  author={Xu, Zheng and Abraham, Jacob},
  booktitle={2019 IEEE International Test Conference (ITC)},
  pages={1--10},
  year={2019},
  organization={IEEE}
}

@inproceedings{ozen2019sanity,
  title={Sanity-check: Boosting the reliability of safety-critical deep neural network applications},
  author={Ozen, Elbruz and Orailoglu, Alex},
  booktitle={2019 IEEE 28th Asian Test Symposium (ATS)},
  pages={7--75},
  year={2019},
  organization={IEEE}
}

@inproceedings{veronesi2024cross,
  title={Cross-layer reliability analysis of nvdla accelerators: Exploring the configuration space},
  author={Veronesi, Alessandro and Nazzari, Alessandro and Passarello, Dario and Krstic, Milos and Favalli, Michele and Cassano, Luca and Miele, Antonio and Bertozzi, Davide and Bolchini, Cristiana},
  booktitle={Proceedings of the 29th IEEE European Test Symposium 2024},
  year={2024}
}

@inproceedings{he2020fidelity,
  title={Fidelity: Efficient resilience analysis framework for deep learning accelerators},
  author={He, Yi and Balaprakash, Prasanna and Li, Yanjing},
  booktitle={2020 53rd Annual IEEE/ACM International Symposium on Microarchitecture (MICRO)},
  pages={270--281},
  year={2020},
  organization={IEEE}
}
\end{document}